\documentclass[useAMS,usenatbib]{mn2e}

\usepackage{aas_macros}

\usepackage{amsmath}
\usepackage{amssymb}
\usepackage{color}
\usepackage{array}
\usepackage{hhline}
\usepackage{graphicx}
\usepackage{float}

\usepackage{booktabs}

\newcommand{\merf}{\mathrm{erf}}
\newcommand{\me}{\mathrm{e}}
\newcommand{\md}{\mathrm{d}}
\renewcommand\Re{\operatorname{\mathfrak{R}}}

\newcommand\rqqpaper[0]{Jackson, et. al. (2015, in prep.)}

\begin{document}
\title{On the use of shapelets in modelling resolved, gravitationally lensed images}
\author[Tagore \& Jackson]{Amitpal S. Tagore and Neal Jackson \\
University of Manchester, School of Physics \& Astronomy, Jodrell Bank Centre for Astrophysics,
Manchester, UK}
\maketitle

\begin{abstract}

Lens modeling of resolved image data has advanced rapidly over the past two decades.
More recently pixel-based approaches, wherein the source is reconstructed on an irregular or adaptive grid, have become popular.
Generally, the source reconstruction takes place in a Bayesian framework and is guided by a set of sensible priors.
We discuss the integration of a shapelets-based method into a Bayesian framework and quantify the required regularization.
In such approaches, the source is reconstructed analytically, using a subset of a complete and orthonormal set of basis functions, known as shapelets.
To calculate the flux in an image plane pixel, the pixel is split into two or more triangles (depending on the local magnification), and each shapelet basis function is integrated over the source plane.
Source regularization (enforcement of priors on the source) can also be performed analytically.
This approach greatly reduces the number of source parameters from the thousands to hundreds and results in a posterior probability distribution that is much less noisy than pixel-based approaches.

\bigskip
\end{abstract}


\section{Introduction}

Since the discovery of the first gravitationally lensed quasars, astronomers have used strong gravitational lensing to constrain cosmological parameters and explore distant quasars \citep[see the review by][]{saasfee}.
Because quasars are typically unresolved, the positions, fluxes, and time delays of the observed images can be used to constrain the lens model.
Galaxies, on the other hand, are extended objects, and there is quite a lot of information buried in how the surface brightness of the images varies pixel to pixel.

Early attempts to model the source's (the galaxy's) light profile assumed elliptical symmetry and analyzed isophotal shapes \citep[e.g.,][]{blandfordEring} or peak surface brightness curves \citep[e.g.,][]{kochanekEring} in Einstein rings.
More general strategies used an iterative, two-loop method to solve for the best lens and source models \citep{wallington1996,koopmans:cdmsubstructure}.
As this can be time-consuming, \citet{citewarrendyesemilinear} introduced a semilinear process in which, for a given lens model, the source could be solved for analytically, greatly speeding up the process.

More recent methods, however, reconstruct the source on an irregular, pixelated grid \citep{dyegrid,vegettigrid}.
Doing so places tighter constraints on lens models and can allow substructure to be detected as well \citep[e.g.,][]{suyu:substr, vegetticlone, vegetti:dark, vegetti:dark2}.
Additionally, there has been an increased interest in exploring the structure of the galaxies being lensed, and it is possible to recover small-scale features in the de-lensed source \citep[e.g.,][]{sourceplanescience,sourceplanescience2,vegettisdp81}.

However, pixel-based reconstruction algorithms require a large number of source parameters, which can reach into the thousands.
Even on modern computers, it can take weeks to months to analyze a single gravitationally lensed system.
Recently, \citet{birrershapelets} introduced a shapelets-based method, wherein the source is reconstructed analytically.
Shapelets are a family of functions that are mathematically complete and orthonormal, and the source's surface brightness is expanded using a a finite subset of the shapelets.
As \citet{refregier_shapeletsI} have demonstrated, the Hermite polynomials can be used to construct shapelets that can reproduce galaxy shapes using a small number of shapelets.
In practice, only a few hundred source parameters are needed to account for the surface brightness distribution seen in the observed images.

Here, we extend the analysis by formulating the shapelets-based source reconstruction in a Bayesian framework.
In order to stabilize the source reconstruction, we place priors on the source's surface brightness (i.e., regularizing the source) and optimize the strength of the regularization by maximizing the Bayesian evidence.
Additionally, we present an image-plane tiling algorithm and integrate the shapelets over the source plane, as would occur naturally during an observation.
We then compare results from the shapelets-based method to existing pixel-based methods.

\section{Shapelets}
\label{sec:shapelets}

\subsection{One-dimensional basis functions}

In order to describe a galaxy using shapelets, we adopt the basis functions developed in \citet{refregier_shapeletsI} and \citet{massey_shapeletsII}.
Here, only the required background information to keep this work self-contained is given.
For a more complete explanation, please see the aforementioned papers.

From \citet{refregier_shapeletsI} the one-dimensional, dimensionless, Cartesian basis functions are given by 
\begin{equation}
b_n(x) \equiv \big( 2^n \pi^\frac{1}{2} n! \big) ^ \frac{1}{2} \, H_n(x)\, \me^{-\frac{x^2}{2}},
\end{equation}
where $H_n$ is the $n^{th}$ order Hermite polynomial.
For describing galaxy light profiles, the basis functions are scaled to obtain
\begin{equation}
\phi_n(x) \equiv \beta^{-\frac{1}{2}}\,\phi_n(\beta^{-1}x),
\end{equation}
where $\beta$ is the characteristic length scale of the galaxy.
Both the dimensional and dimensionless basis functions are, separately, complete and orthonormal.
Thus a galaxy, no matter how complex, can be described using these shapelets.

\subsection{Two-dimensional, Cartesian shapelets}

Two-dimensional, Cartesian basis functions can be constructed from the tensor product of the one-dimensional functions.
We have,
\begin{equation}
\mathbf{\Phi}_{n_x,n_y}(x,y) \equiv \beta^{-1}\, \phi_{n_x}(\beta^{-1}x) \,\otimes\, \phi_{n_y}(\beta^{-1}y).
\end{equation}

For a galaxy, its surface brightness distribution, $S(x,y)$, can be expanded as
\begin{equation}
S(x,y) = \sum_{n_x=0}^\infty \sum_{n_y=0}^\infty c_{n_x,n_y} \mathbf{\Phi}_{n_x,n_y}(x,y),
\end{equation}
and the shapelet coefficients, $c_{n_x,n_y}$, are given by the overlap integral
\begin{equation}
c_{n_x,n_y} = \int\int \md x\,\md y\,S(x,y)\,\mathbf{\Phi}_{n_x,n_y}(x,y).
\end{equation}

\subsection{Two-dimensional, polar shapelets}

We can also describe the source using a polar coordinate system, which is a more natural choice for galaxies.
From \citet{massey_shapeletsII} the two-dimensional basis functions in this case are given by
\begin{equation}
\begin{aligned}
\chi_{n,m}(r,\theta) = &\frac{(-1)^\frac{n-\mid m\mid}{2}}{\beta^{\mid m\mid+1}} \bigg( \frac{(\frac{n-\mid m\mid}{2})!}{\pi(\frac{n+\mid m\mid}{2})!} \bigg)^\frac{1}{2} r^{\mid m\mid} \times\
\\
&L^{\mid m\mid}_\frac{n-\mid m\mid}{2} \bigg(\frac{r}{\beta}\bigg)^2 \me^{-\frac{r^2}{2\beta^2}} \me^{-im\theta},
\end{aligned}
\end{equation}
where $r$ and $\theta$ are the radial and angular coordinates, respectively, $\beta$ is (as before) the characteristic scale length, and $L^q_p(x)$ are the associated Laguerre polynomials.
The order of the basis function is determined by $n$ and $m$.
$n$ can be any non-negative integer. $m$ can take on values between $-n$ and $n$ but must increase in steps of two.
That is, $m=-n, -n+2,...,n-2,n$.

As before, the polar shapelets are complete and orthonormal, and we can expand the source's surface brightness as
\begin{equation}
S(r,\theta) = \sum_{n=0}^\infty\sum_{m=-n}^n c_{n,m} \, \chi_{n,m}(r,\theta;\beta),
\end{equation}
and the shapelet coefficients are given by the overlap integral
\begin{equation}
c_{n,m} = \iint_{\Re} r \md r \,\md\theta \,S(r,\theta) \,\chi_{n,m}(r,\theta;\beta)
\end{equation}
As mentioned, the summation over $m$ assumes $m$ increases in steps of two.

Unlike the Cartesian case, the polar basis functions and their shapelet coefficients are complex.
The magnitude of a given complex shapelet coefficient determines how strongly that shapelet contributes to the surface brightness; the phase determines the angle of orientation.
We note that although the complex polar coefficients contain twice as many free parameters than do an equivalent number of Cartesian coefficients, the polar coefficients are not all independent.
Specifically, $c_{n,m}=c_{n,-m}^*$, where the asterisk denotes complex conjugation.
This relation implies that there is the same amount of information encoded in both the Cartesian and polar coefficients.

\subsection{Practical issues}

There are several free parameters when performing a shapelets reconstruction.
\citet{massey_shapeletsII} present a set of criteria and an iterative approach for optimizing the shapelet center and scale, as well as the maximum order of the shapelets basis functions.
Because this approach is time-consuming when searching the lens model parameter space, we instead fix all three of these parameters.
The maximum order of the shapelets is fixed but can be adjusted later if necessary.
\citet{birrershapelets} also make these simplifying assumptions and apply their method to the gravitational lens RXJ1131-1231.
In \S\ref{sec:lo} and \S\ref{sec:reg}, we discuss issues related to integrating the shapelets over the source plane and fixing the maximum shapelets order.
We also explore the effects the shapelets order can have on the reconstruction in \S\ref{sec:test_shapelets}.

The shapelets center and scale, on the other hand, vary from one lens model to the next.
To estimate the center, we ray-trace image pixels with surface brightness values greater than three times the noise level.
The surface brightness weighted positions of the subset of image pixels are used to fix the shapelets center.
Similarly, the surface brightness weighted distances of the ray-traced pixels to the center are used to fix the scale.\footnote{If the noise is not uniform across the image, inverse noise weighting can be used as well.}

In order to easily integrate over the the shapelets basis functions (see \S\ref{sec:lo}), we choose to use the Cartesian shapelets when performing the reconstruction.
However, as noted in \citet{massey_shapeletsII}, these can be later converted to polar shapelets, which are useful for obtaining analytic functions for properties of the source, such as the azimuthally averaged radial light profile.

\section{Source reconstruction methodology}
\label{sec:bayesian}

With a few modifications, shapelets can enter into the established Bayesian frameworks for analyzing resolved gravitationally lensed images using pixelated source grids.
In these frameworks, the ideal, reconstructed source reproduces the data, while not violating any priors placed on it.
In this section we first briefly review the formal Bayesian framework, which has been discussed in detail by \citet{citewarrendyesemilinear}, \citet{suyureg}, \citet{vegettigrid}, and \citet{citebrewer2006}.
Then, we show how shapelets can replace a pixelated source grid.
We note that lens model ranking is not discussed here but is addressed in the aforementioned papers.

\subsection{Bayesian framework}

In the absence of scattering or absorption of light, lensing conserves surface brightness.
We can therefore relate the source and lensed images as 
\begin{equation}
\label{lensequation1}
\mathbf{d}=\mathbf{L}\mathbf{s} + \mathbf{n},
\end{equation}
where $\mathbf{L}$ is a linear ``lensing operator'' that include gravitational, atmospheric, and instrumental effects.
$\mathbf{s}$ and $\mathbf{d}$ are vectors containing the surface brightness values in the source plane and image plane, respectively, and $\mathbf{n}$ is the noise present in the data.

To reconstruct the source, a penalty function, $P$, is introduced:
\begin{equation}
P = G + \lambda R,
\end{equation}
where $G$ quantifies the mismatch between the data and the model, $R$ quantifies how much the source violates priors placed upon it, and $\lambda$ is the regularization strength, which determines how strongly the priors are enforced.
Specifically,
\begin{equation}
G = \frac{1}{2}(\mathbf{Ls-d})^\top \mathbf{C}_\text{d}^{-1}(\mathbf{Ls-d}),
\end{equation}
where $\mathbf{C}_\text{d}$ is the noise covariance matrix of the data, and 
\begin{equation}
R = \frac{1}{2}(\mathbf{H}\mathbf{s})^\top\mathbf{H}\mathbf{s},
\end{equation}
where $\mathbf{H}$ encodes the priors placed on the source.
For example, $\mathbf{H}$ could act on a source vector $\mathbf{s}$ to produce a vector of derivatives of the source's surface brightness.
$R$ would then be the sum of the squares of the derivatives of the source's surface brightness, within a factor of two.

Minimizing $P$ leads to an equation for the reconstructed source
\begin{equation}
\mathbf{s}_r = \mathbf{F}^{-1}\mathbf{D},
\end{equation}
where $\mathbf{F}=\mathbf{L}^\top\mathbf{C}_\text{d}^{-1}\mathbf{L} + \lambda\mathbf{H}^\top\mathbf{H}$ and $\mathbf{D}=\mathbf{L}^\top\mathbf{C}_\text{d}^{-1}\mathbf{d}$.

We note that it is also possible to non-linearly solve for the optimal regularization strength by minimizing $P$.
For details of the derivation and optimization, see \citet{suyureg}.

\subsection{Constructing the lensing operator}
\label{sec:lo}

The primary difference between pixel-based and shapelets-based source reconstruction methods lies in the construction of the lensing operator.
In pixel-based methods \citep[see, e.g.,][]{vegettigrid}, each image-plane pixel is ray-traced to the source plane.
The irregular, pixelated source grid is triangulated, and the three source pixels that enclose the ray-traced data pixel are identified.
Bilinear weights are placed on these three source pixels so that the total surface brightness matches the data pixel's surface brightness. 
In this way, the lensing operator is constructed, and there are at most three entries per row.\footnote{Before convolution with the instrumental PSF or any other kernels.}

When using shapelets, there is no source grid, and the source vector does not contain surface brightness values.
Instead, the source vector contains the strengths of the basis functions, and the position along the source vector determines the order of the basis function.
In order to create the lensing operator, each data pixel is ray-traced to the source plane, as before.
The location of a ray-traced pixel determines how strongly each basis function contributes (i.e., the basis function evaluated at the location) to the surface brightness of that particular data pixel.
Thus, the lensing operator contains no empty elements.
However, this does not necessarily imply that the shapelets-based approach is slower or more computationally expensive, as the dimensions of the lensing operator are much fewer than in the pixel-based case.

In the absence of a gravitational lens, there is a minimum (set by the pixel size) and maximum (set by the image dimensions) scale over which information can be obtained \citep{refregier_shapeletsI}. 
These extremes will constrain the maximum order of the basis functions and shapelets scale that should be used in reconstructing the image.
However, because gravitational lensing magnifies some regions of the image more than others, determining the maximum shapelet order and scale is not straightforward.
Whereas image pixels in highly magnified regions may require higher-order shapelets, pixels in lower magnification regions may, essentially, randomly sample the highly oscillatory shapelet.

In order to combat this issue, when calculating the surface brightness in an image pixel, we integrate each shapelet over the ray-traced region in the source plane.
Because highly oscillatory shapelets will tend to average to zero over large regions, we fix the maximum order of the shapelets at some reasonable value (i.e., $20\times 20-50\times 50$).
The shapelet integration is complicated by the fact that a square image pixel will not map to a square region in the source plane, and it may, in fact, overlap on itself.
We thus split each image pixel into $N_\textnormal{split}$ right triangles, because triangles will always map to triangles and are more easily handled.\footnote{This image-plane tiling scheme is analogous to the one presented in \citet{citelensmodel}, although differences are discussed in the text.}
For a particular image plane pixel, each of the $N_\textnormal{split}$ integrals will give the average surface brightness over that triangle.
The average surface brightness over the image plane pixel $p$ is then given by
\begin{equation}
  S_p = N_\textnormal{split}^{-1} \sum\limits_{t=1}^{N_\textnormal{split}}A_t^{-1}\sum\limits_{n_x}\sum\limits_{n_y}\int\int\md x\md y\Phi_{n_x,n_y}(x,y),
\end{equation}
where $A_t$ is the area of the ray-traced triangle.

$N_\textnormal{split}$ can vary from one image pixel to the next.
We choose the value based on the local magnification, so that, for a particular pixel,
\begin{equation}
N_\textnormal{split} = \lfloor2\mu\rfloor,
\end{equation}
where $\mu$ is the magnification at the center of the pixel and the half square brackets denote the floor function.
This prescription allows $N_\textnormal{split}$ to follow the magnification relatively smoothly across the image plane.
However, we allow for minimum (two or greater) and maximum values of $N_\textnormal{split}$ to be enforced.

\begin{figure}
  \centering
  \includegraphics[width=0.45\textwidth]{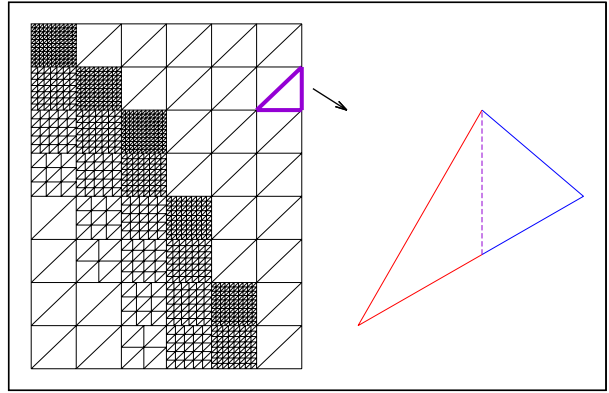}
  \caption{\small
    Schematic diagram illustrating the shapelets integration. First, the image pixels are split into right triangles, where the number of splittings depends on the local magnification (or can be fixed to a constant value). These triangles are then mapped back to the source plane. Focusing on one ray-traced triangle, it is further divided into two smaller triangles, such that the two share the vertex with the median $x$-coordinate, and the line segment that splits the triangle has infinite slope. Then, two integrations are performed. Each integration has fixed upper and lower limits over the $x$-coordinate. The limits of the integral over the $y$-coordinate are, however, functions of the $x$-coordinate and are denoted by the solid, blue and red lines.}
  \label{fig:triangle-splitting}
\end{figure}

Additionally, a ray-traced triangle will not remain a right triangle, and the integration will be performed in two steps as follows.
We choose to fix the lower and upper limits of integration in the $x$ direction.
Subsequently, the lower and upper limits of integration in the $y$ direction will be a function of the $x$-coordinate.
We therefore split each ray-traced triangle into two smaller triangles, such that both triangles share the vertex with the median $x$-coordinate, and the line segment that splits the triangle has infinite slope.
A schematic diagram is shown in Fig~\ref{fig:triangle-splitting}.

There is no closed-form solution for integrating the triangles in the source plane.
We instead present a series of recursion relation which can be used to quickly calculate the integrals.
For fixed limits of integration, \citet{massey_shapeletsII} show that the one-dimensional integral over the shapelets is given by
\begin{equation}
  \begin{aligned}
  I_n(a,b) &\equiv \int\limits_a^b\md x' \phi_n(x') \\
  &= \sqrt{\frac{n-1}{n}}\;I_{n-2}(a,b) - \beta\sqrt{\frac{2}{n}}\;\phi_{n-1}(x')\bigg|_{x'=a}^b
  \end{aligned}
\end{equation}

We seek to evaluate the two-dimensional integral given by
\begin{equation}
\begin{aligned}
  I_{n_x,n_y} &\equiv \int\limits_c^d\md x\int\limits_{m_1x+b_1}^{m_2x+b_2}\md y\;\phi_{n_x}(x)\;\phi_{n_y}(y) \\ 
  &= \int\limits_c^d\md x \; \phi_{n_x}(x)\; I_{n_y}(m_1 x+b_1,m_2 x+b_2) \\
  &= \sqrt{\frac{n_y-1}{n_y}}I_{n_x,n_y-2} + J_{n_x,n_y-1},
\end{aligned}
\end{equation}
where we have defined
\begin{equation}
J_{n_x,n_y}\equiv -\beta\sqrt{\frac{2}{n_y+1}} \int\limits_c^d \md x \; \phi_{n_x}(x)\phi_{n_y}(x')\bigg|_{x'=m_1 x+b_1}^{m_2 x+b_2}.
\end{equation}
Note that $x'$ is evaluated as a function of the dummy variable $x$ over which the integral is evaluated.

Before calculating $J_{n_x,n_y}$, we first present recursion relations for the $I_{n_x,n_y}$ with $n_y=0,1$.
From hereon out in order to keep the equations tractable, we will no longer explicitly write the evaluation of dummy variables at the limits of the integrals.
For example, we will denote an expression of the form $(m x +b)\vert_{m=m_1;b=b_1}^{m_2; b_2}\vert_{x=c}^{d} = ((m_2 x+b_2)-(m_1 x+b_1))\vert_{x=c}^{d} = ((m_2 d+b_2)-(m_1 d+b_1))-((m_2 c+b_2)-(m_1 c+b_1))$ simply as $(m x+b)$.
Thus, the remainder of the equations in this section may actually consist of up to four times as many terms as are explicitly written.
Using this notation, we have

\begin{equation}
\begin{aligned}
  I_{n_x,1} &= \bigg[(m^2+1)\sqrt{n_x}\bigg]^{-1}\;\times \\
  &\bigg[2\beta^\frac{3}{2}[\pi^\frac{1}{4}]^{-1}\phi_{n_x-1}(x)\me^{-(m x+b)^2/(2\beta^2)} \\
  &-\sqrt{2} \frac{m b}{\beta} I_{n_x-1,1} \\
  &-(m^2-1)\sqrt{n_x-1} I_{n_x-2,1}\bigg]
  \end{aligned}
\end{equation}

and

\begin{equation}
\begin{aligned}
  \sqrt{n_x}\;I_{n_x,0} = &-\pi^\frac{1}{4}\beta^\frac{3}{2}\merf\bigg(\frac{m x+b}{\beta\sqrt{2}}\bigg)\phi_{n_x-1}(x) \\
&-m I_{n_x-1,1} \\
&+ \sqrt{n_x-1}I_{n_x-2,0}
  \end{aligned}
\end{equation}

The starting points for these relations are given by 
\begin{equation}
I_{0,1} = - \frac{\beta}{\sqrt{1+m^2}}\merf\bigg(\frac{x+m(b+m x)}{\sqrt{2\beta^2(1+m^2)}}\bigg)\me^{-b^2/(2\beta^2(1+m^2))},
\end{equation}
\begin{equation}
\begin{aligned}
I_{1,0} = &-\beta\merf\bigg(\frac{b + m x}{\sqrt{2}\beta}\bigg)\me^{-x^2/(2 \beta^2)} \\
&+ \frac{\beta m}{\sqrt{1 + m^2}} \merf\bigg(\frac{x + m (b + m x)}{\sqrt{2\beta^2(1 + m^2)}}\bigg) \me^{-b^2/(2\beta^2(1 + m^2))},
\end{aligned}
\end{equation}
and
\begin{equation}
\begin{aligned}
I_{1,1} = &\me^{-(b^2+2 b m x+(1+m^2)x^2)/(2\beta^2)}(1+m^2)^{-3/2}\pi^{-1/2} \times \\
&\bigg[2\beta\sqrt{1+m^2} + \sqrt{2\pi}b m \me^{(x+m(b+m x))^2/(2\beta^2(1+m^2))} \times \\
&\merf\bigg(\frac{x+m(b+m x)}{\sqrt{2\beta^2(1+m^2)}}\bigg)\bigg]
\end{aligned}
\end{equation}

To our knowledge, there does not exist a closed form for $I_{0,0}$.
It can be computed numerically, or by approximating the error function as the product of an exponential function and a truncated power series expansion, $I_{0,0}$ can be computed analytically to high accuracy. See the appendix for details.

It remains to evaluate the integrals $J_{n_x,n_y}$.
Again, we present recursion relations to compute these quickly.
\begin{equation}
\begin{aligned}
J_{n_x,n_y} &= \bigg[(m^2+1)\sqrt{n_x}\bigg]^{-1}\;\times \\
&\bigg[\frac{2\beta^2}{\sqrt{n_y+1}} \phi_{n_x-1}(x)\phi_{n_y}(m x+b) \\
&-b m \beta^{-1}\sqrt{2} J_{n_x-1,n_y} \\
&-(m^2-1)\sqrt{n_x-1} J_{n_x-2,n_y} \\
&+\frac{2 m n_y}{\sqrt{n_y+1}} J_{n_x-1,n_y-1}\bigg]
\end{aligned}
\end{equation}

All the $J_{n_x,n_y}$ integrals are fully specified with the above relation and the following:
\begin{equation}
\begin{aligned}
J_{0,n_y} = &\bigg[(m^2+1)\sqrt{n_y+1}\bigg]^{-1}\;\times \\
&\bigg[\frac{2 m \beta^\frac{3}{2}}{\pi^\frac{1}{4}\sqrt{n_y}}\phi_{n_y-1}(m x+b)\me^{-x^2/(2\beta^2)} \\
&+\sqrt{2} b/\beta J_{0,n_y-1} \\
&+(m^2-1)(n_y-1)/\sqrt{n_y}J_{0,n_y-2}\bigg],
\end{aligned}
\end{equation}
\begin{equation}
\begin{aligned}
J_{1,n_y} = &\frac{\sqrt{n_y+2}}{m} J_{0,n_y+1} \\
&+ \frac{n_y}{m\sqrt{n_y+1}} J_{0,n_y-1} - \frac{\sqrt{2}b}{m\beta} J_{0,n_y},
\end{aligned}
\end{equation}
\begin{equation}
J_{n_x,0} = I_{n_x,1},
\end{equation}
and
\begin{equation}
  \begin{aligned}
&J_{0,1} = \bigg[m^2+1\bigg]^{-1} \times \bigg[ \sqrt{\frac{2}{\pi}}m\beta\me^{-((m x+b)^2+x^2)/(2\beta^2)} \\
&- (m^2+1)^{-\frac{1}{2}}b \me^{-b^2/(2(m^2+1)\beta^2)}\merf\bigg(\frac{(m^2+1)x+m b}{\sqrt{2(m^2+1)}\beta}\bigg)\bigg]
\end{aligned}
\end{equation}

\subsection{Regularizing the source}
\label{sec:reg}

Just as with pixel-based source reconstruction methods, regularization is necessary, especially since the maximum shapelet order is fixed and not set by the data.
Several forms of regularization have become popular in the literature \citep[see, e.g.,][]{suyureg}.
We present formulae for calculating the zeroeth, first, and second derivatives of the surface brightess and for calculating the RMS size of the source.
Shapelet indices with a hat above them (e.g., $\hat{n}_x$) denote the maximum order of the shapelet along the specified coordinate.

Zeroth order regularization penalizes sources with large surface brightness values and is given by the integral over all space of the square of the surface brightness:
\begin{equation}
R_0 \equiv \frac{1}{2}\sum\limits_\mathbf{n}\int\limits_{-\infty}^{\infty}\int\limits_{-\infty}^{\infty}\md x\md y \bigg(c_\mathbf{n}\Phi_\mathbf{n}(\mathbf{x})\bigg)^2 = \frac{1}{2}\sum\limits_\mathbf{n} c_\mathbf{n}^2.
\end{equation}

If we instead want to minimize the change in surface brightness, then a possible choice for gradient regularization is given by the integral over all space of the square of the magnitude of the gradient of the surface brightness:
\begin{equation}
  \begin{aligned}
    R_1 &\equiv \frac{1}{2}\sum\limits_\mathbf{n}\int\limits_{-\infty}^{\infty}\int\limits_{-\infty}^{\infty}\md x\md y \bigg(c_\mathbf{n}\vec{\nabla}\Phi_\mathbf{n}(\mathbf{x})\bigg)^2 \\
    &= \frac{1}{4\beta^2}\sum\limits_{n_y} \bigg[\Big(c_{1,n_y}\Big)^2 + \sum\limits_{n_x=\hat{n}_x}^{\hat{n}_x+1} \Big(\sqrt{n_x}c_{n_x-1,n_y}\Big)^2 \\
    &+ \sum\limits_{n_x=1}^{\hat{n}_x-1} \Big(\sqrt{n_x}c_{n_x-1,n_y}-\sqrt{n_x+1}c_{n_x+1,n_y}\Big)^2\bigg] \\
    &+ \frac{1}{4\beta^2}\sum\limits_{n_x} \bigg[\Big(c_{n_x,1}\Big)^2 + \sum\limits_{n_y=\hat{n}_y}^{\hat{n}_y+1} \Big(\sqrt{n_y}c_{n_x,n_y-1}\Big)^2 \\
    &+ \sum\limits_{n_y=1}^{\hat{n}_y-1} \Big(\sqrt{n_y}c_{n_x,n_y-1}-\sqrt{n_y+1}c_{n_x,n_y+1}\Big)^2\bigg]
  \end{aligned}
\end{equation}

If we want to ensure that the surface brightness is smoother over even larger scales, then a possible choice for curvature regularization is given by the integral over all space of the square of the Laplacian of the surface brightness:
\begin{equation}
  \begin{aligned}
    R_2 &\equiv \frac{1}{2}\sum\limits_\mathbf{n}\int\limits_{-\infty}^{\infty}\int\limits_{-\infty}^{\infty}\md x\md y \bigg(c_\mathbf{n}\nabla^2\Phi_\mathbf{n}(\mathbf{x})\bigg)^2 \\
    &=\frac{1}{8\beta^4}\Bigg\{\sum\limits_{n_x=0}^{1}\bigg[\sum\limits_{n_y=0}^{1}(K_{1x}(n_x, n_y) + K_{1y}(n_x, n_y))^2 \\
      &+\sum\limits_{n_y=2}^{\hat{n}_y-2}(K_{1x}(n_x,n_y) + K_{2y}(n_x,n_y))^2 \\
      &+\sum\limits_{n_y=\hat{n}_y-1}^{\hat{n}_y}(K_{1x}(n_x,n_y) + K_{3y}(n_x,n_y))^2\bigg] \\
      &+\sum\limits_{n_x=2}^{\hat{n}_x-2}\bigg[\sum\limits_{n_y=0}^{1}(K_{2x}(n_x,n_y) + K_{1y}(n_x,n_y))^2 \\
      &+\sum\limits_{n_y=2}^{\hat{n}_y-2}(K_{2x}(n_x,n_y) + K_{2y}(n_x,n_y))^2 \\
      &+\sum\limits_{n_y=\hat{n}_y-1}^{\hat{n}_y}(K_{2x}(n_x,n_y) + K_{3y}(n_x,n_y))^2\bigg] \\
      &+\sum\limits_{n_x=\hat{n}_x-1}^{\hat{n}_x}\bigg[\sum\limits_{n_y=0}^{1}(K_{3x}(n_x,n_y) + K_{1y}(n_x,n_y))^2 \\
      &+\sum\limits_{n_y=2}^{\hat{n}_y-2}(K_{3x}(n_x,n_y) + K_{2y}(n_x,n_y))^2 \\
      &+\sum\limits_{n_y=\hat{n}_y-1}^{\hat{n}_y}(K_{3x}(n_x,n_y) + K_{3y}(n_x,n_y))^2\bigg] \\
      &+\sum\limits_{n_y=0}^{\hat{n}_y}\sum\limits_{n_x=\hat{n}_x+1}^{\hat{n}_x+2}(\sqrt{n_x (n_x - 1)} c_{n_x-2,n_y})^2 \\
      &+\sum\limits_{n_x=0}^{\hat{n}_x}\sum\limits_{n_y=\hat{n}_y+1}^{\hat{n}_y+2}(\sqrt{n_y (n_y - 1)} c_{n_x,n_y-2})^2\Bigg\},
  \end{aligned}
\end{equation}
where
\begin{equation}
  \begin{aligned}
    K_{1x}(n_x,n_y) &= \sqrt{(n_x + 1) (n_x + 2)} c_{n_x + 2, n_y} - (2 n_x + 1) c_{n_x, n_y} \\
    K_{1y}(n_x,n_y) &= \sqrt{(n_y + 1) (n_y + 2)} c_{n_x, n_y + 2} - (2 n_y + 1) c_{n_x, n_y} \\
    K_{2x}(n_x,n_y) &= \sqrt{(n_x + 1) (n_x + 2)} c_{n_x + 2, n_y} \\
    &- (2 n_x + 1) c_{n_x, n_y} + \sqrt{n_x (n_x - 1)} c_{n_x - 2, n_y} \\
    K_{2y}(n_x,n_y) &= \sqrt{(n_y + 1) (n_y + 2)} c_{n_x, n_y + 2} \\
    &- (2 n_y + 1) c_{n_x, n_y} + \sqrt{n_y (n_y - 1)} c_{n_x, n_y - 2} \\
    K_{3x}(n_x,n_y) &= -(2 n_x + 1) c_{n_x, n_y} + \sqrt{n_x (n_x - 1)} c_{n_x - 2, n_y} \\
    K_{3y}(n_x,n_y) &= -(2 n_y + 1) c_{n_x, n_y} + \sqrt{n_y (n_y - 1)} c_{n_x, n_y - 2}
  \end{aligned}
\end{equation}

Finally, if the source is known or expected to be compact, we can instead minimize a sort of un-normalized RMS radius, given by
\begin{equation}
  \begin{aligned}
    R_\textnormal{size} &\equiv\frac{1}{2}\sum\limits_\mathbf{n}\int\limits_{-\infty}^{\infty}\int\limits_{-\infty}^{\infty}\md x\md y \bigg(\sqrt{x^2+y^2}c_\mathbf{n}\Phi_\mathbf{n}(\mathbf{x})\bigg)^2 \\
    &=\frac{\beta^2}{4}\sum\limits_{n_y}\bigg[\Big(c_{1,n_y}\Big)^2 + \sum\limits_{n_x=\hat{n}_x}^{\hat{n}_x+1}\Big(\sqrt{n_x}c_{n_x-1,n_y}\Big)^2 \\
    &+ \sum\limits_{n_x=1}^{\hat{n}_x-1}\Big(\sqrt{n_x}c_{n_x-1,n_y}+\sqrt{n_x+1}c_{n_x+1,n_y}\Big)^2\bigg] \\
    &+\frac{\beta^2}{4}\sum\limits_{n_x}\bigg[\Big(c_{n_x,1}\Big)^2 + \sum\limits_{n_y=\hat{n}_y}^{\hat{n}_y+1}\Big(\sqrt{n_y}c_{n_x,n_y-1}\Big)^2 \\
    &+ \sum\limits_{n_y=1}^{\hat{n}_y-1}\Big(\sqrt{n_y}c_{n_x,n_y-1}+\sqrt{n_y+1}c_{n_x,n_y+1}\Big)^2\bigg]
  \end{aligned}
\end{equation}

\section{Testing the shapelets}
\label{sec:test_shapelets}

\subsection{Varying shapelet parameters}

\begin{figure*}
  \centering
  \includegraphics[width=0.95\textwidth]{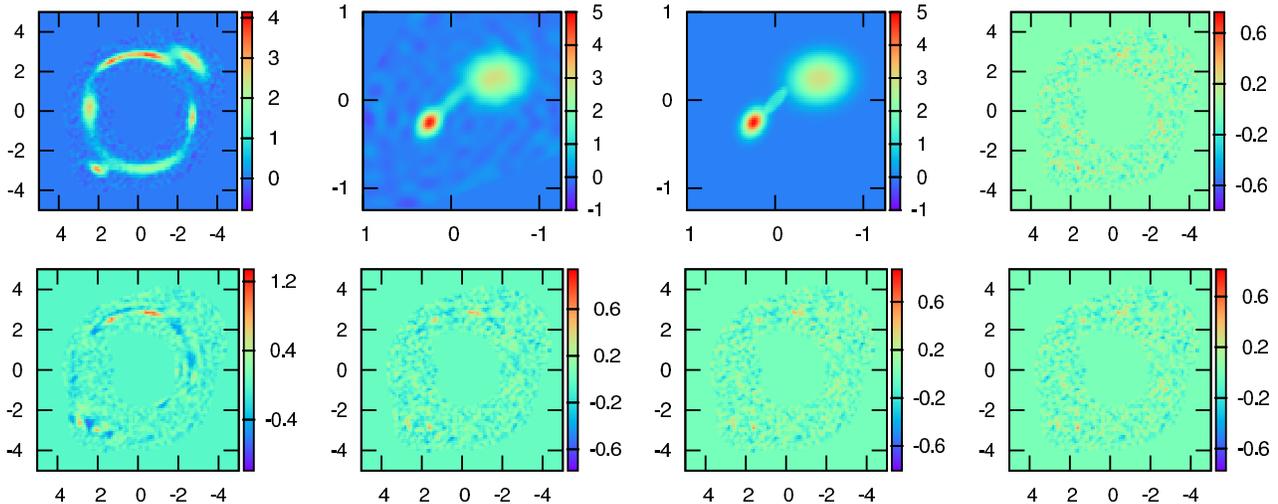}
  \caption{\small
    Shapelets reconstruction for several shapelets configurations. Top row, left to right: data, source reconstruction, true source, model residuals. $50\times 50$ shapelets basis function were used. Bottom row, left to right: Model residuals for runs with $10\times 10$, $20\times 20$, $30\times 30$, $40\times 40$ shapelets basis functions. All runs used a maximum $N_\textnormal{split}$ of $2\times 10^6$ and used optimal gradient regularization. We find that, for this particular mock data, lowering $N_\textnormal{split}$ has no significant effect.}
  \label{fig:merger}
\end{figure*}

Here, we look for effects of varying the maximum shapelet order and the maximum value for $N_\textnormal{split}$ on the source reconstruction.
Using mock data exhibiting complex morphologies, we also test the ability of the shapelets reconstruction to capture both small and large scale features.
The source being lensed is comprised of two merging galaxies, with a bridge connecting the two, and crosses a caustic, so that some regions are doubly imaged, while others are quadruply imaged.
The lens is a singular isothermal ellipsoid with ellipticity of 0.3 and position angle of $60^\circ$, east of north (EoN).
The Einstein radius along the major axis is $3''$, where the pixel scale is 0.05 arcs/pixel.
An elliptical Gaussian PSF has been applied, with a full-width half-maximum (FWHM) of $0.236''$ and $0.165''$ along the major and minor axes, respectively, and a position angle of $-55^\circ$ EoN.
Finally, additive Gaussian noise has been applied, so that the final data has a peak S/N$\approx 21$.
With the lens model fixed at the true parameters, we perform source reconstructions for several different shapelets configurations, shown in Fig.~\ref{fig:merger}.
We find that using $10\times 10$ basis functions is not adequate to capture all of features in the data.
$20\times 20$ basis functions or more do a reasonable job of capturing all features, with the $50\times 50$ case resulting in a reduced $\chi^2 =1$ ($\chi^2$ per pixel).
For this particular mock data, increasing the maximum value of $N_\textnormal{split}$ from two to $2\times 10^6$ does not result in a significant improvement for any of the model residuals shown.

However, we find that $N_\textnormal{split}$ can become important if the source size is reduced and/or moved near a caustic.
In order to explore this further, using the same lens as was used in the preceding mock data analysis, we fix the source to be a circular Gaussian.
The source position is fixed $0.25''$ along the major axis (within the quadruply imaged region), while the FWHM of the source and $N_\textnormal{split}$ are varied.
We then limit the shapelets to $1\times 1$ basis functions (identical to a circular Gaussian) and fix the shapelets center and scale to the true values for each run.
Using this setup, we can attribute any inability of the shapelets integration to recover the true image plane flux to an insufficiently small value of $N_\textnormal{split}$.

\begin{figure}
  \centering
  \includegraphics[width=0.45\textwidth]{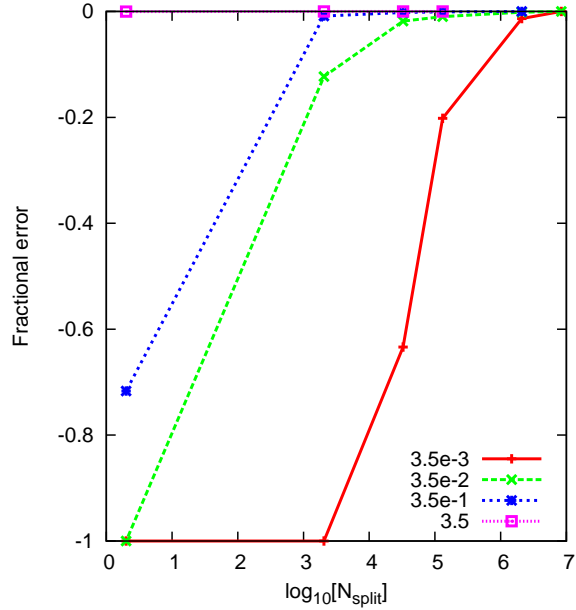}
  \caption{\small
    Fractional error in the flux calculation for several different source sizes as a function of $N_\textnormal{split}$. Different lines represent different source sizes, measured in pixel units at FWHM. The fractional error is computed relative to the flux calculated using the maximum $N_\textnormal{split}$ for that source size. Note: a fractional error of -1 indicates that no flux was recovered in the image plane.}
  \label{fig:nsplit}
\end{figure}

We find that, for larger source sizes, $N_\textnormal{split}$ has little to no effect, but as the source size becomes smaller, a higher $N_\textnormal{split}$ is needed to accurately calculate the flux.
See Fig.~\ref{fig:nsplit}.
For the smallest source size ($3.5\times 10^{-3}$ pixels at FWHM) and highest value of $N_\textnormal{split}$ ($8.4\times 10^6$), the image plane flux of each image calculated using the shapelets method agrees with the results for a point source to within 2\%.\footnote{As the Gaussian source become smaller, the total flux calculated should approach the analytic results for a point source.}
For the next smallest source, ($3.5\times 10^{-2}$ pixels), agreement is to within 0.5\%.
For larger sources, finite source effects become important.
For these reasons, we quote the percent error for each source size relative to the flux calculated using the maximum value of $N_\textnormal{split}$, which is limited by the computational resources required, for that particular source size.

For radio observations of quasars, finite source effects of the emitting region can become significant and distinguishable from a point source model.
In such cases, $N_\textnormal{split}$ can play an important role in accurately recovering image-plane fluxes.
In \rqqpaper{}, we investigate such finite source effects from several radio-quiet quasars.

\subsection{Varying lens model parameters}

\begin{figure}
  \centering
  \begin{tabular}{ c }
  \includegraphics[width=0.45\textwidth]{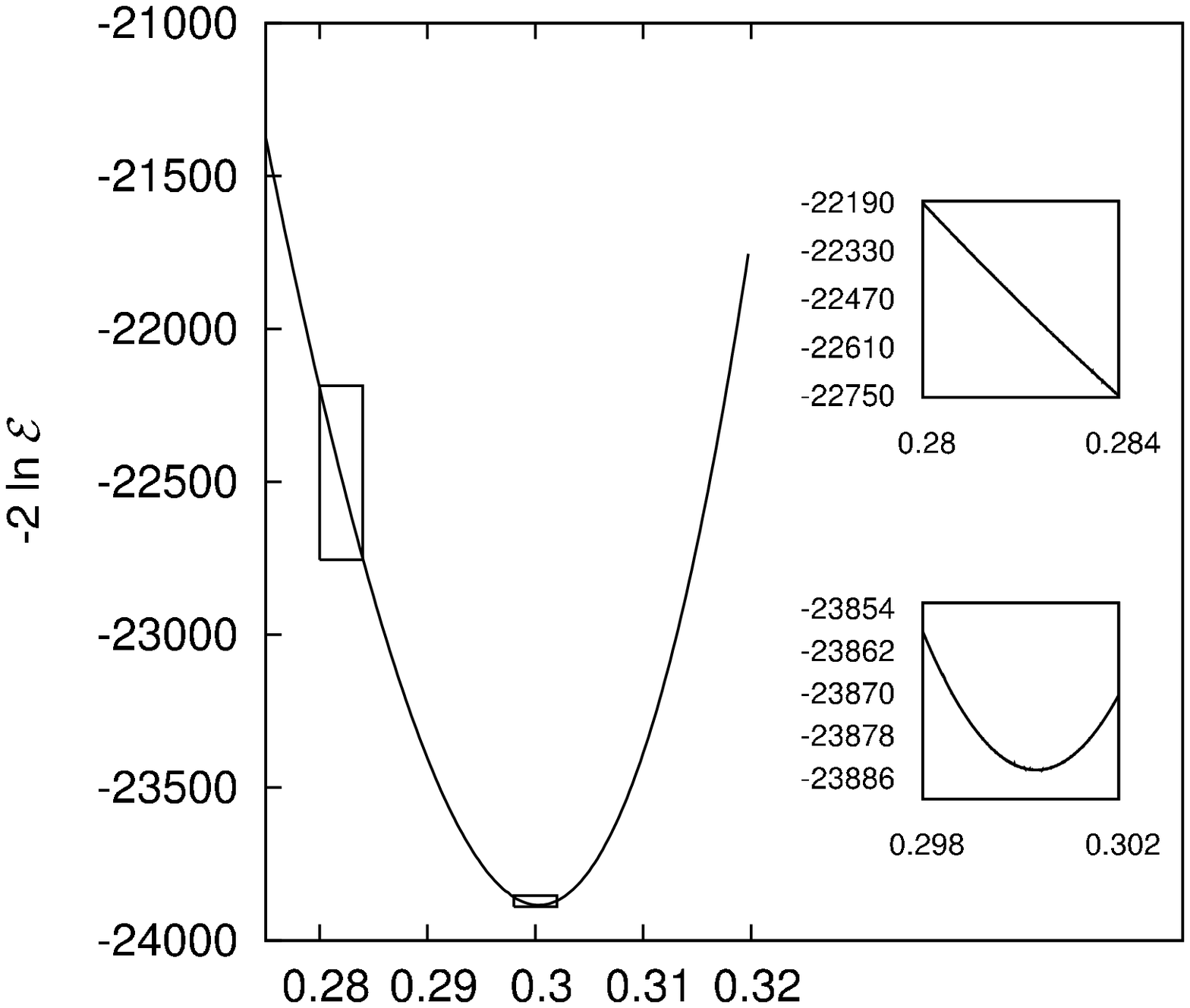} \\
  \includegraphics[width=0.45\textwidth]{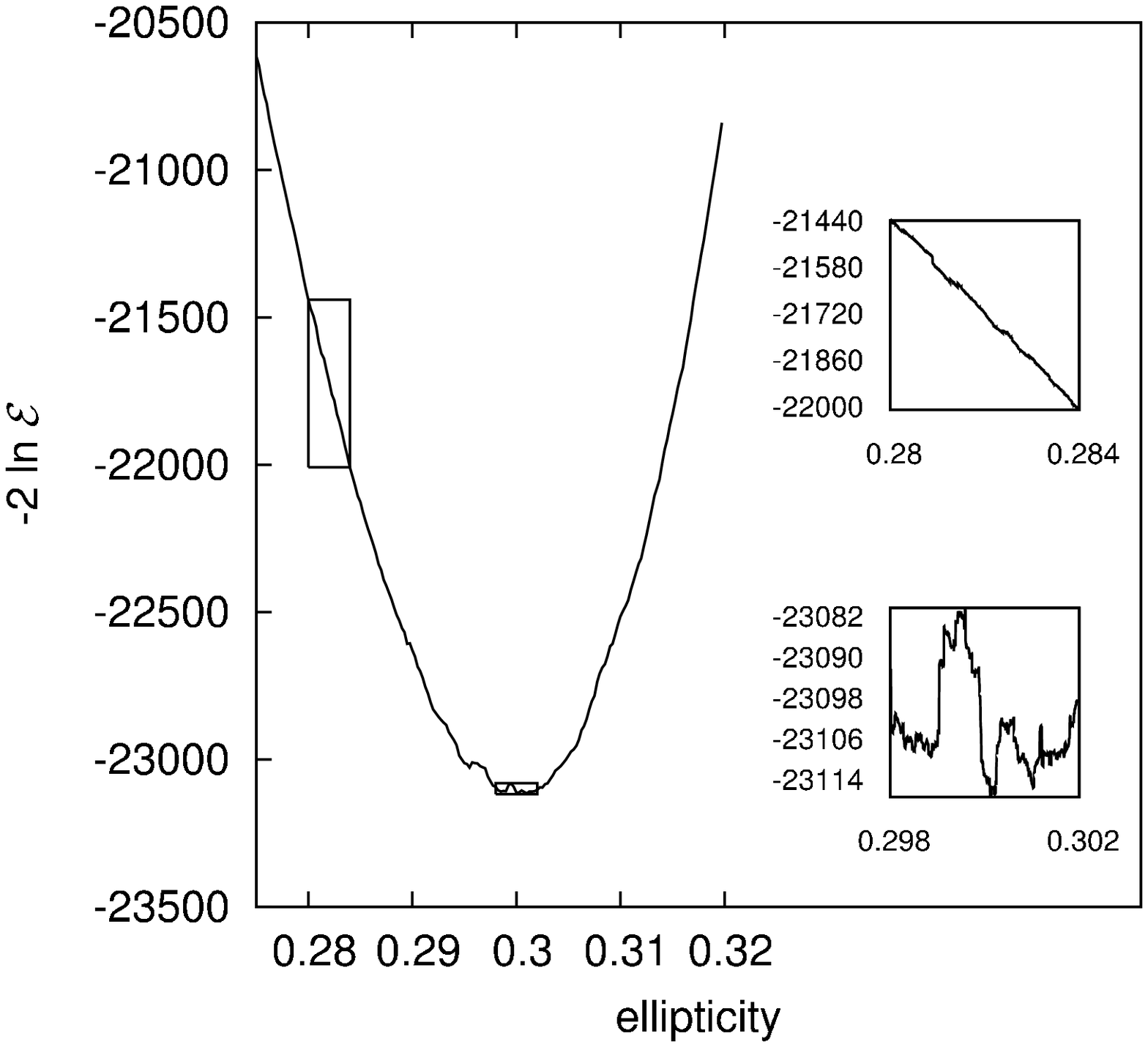}
  \end{tabular}
  \caption{\small
    One dimensional projection of $-2 \ln \mathcal{E}$. All model parameters are fixed at the true values, while ellipticity is varied about its true value of $e=0.3$. Top: Projection using shapelets, with $20\times 20$ shapelets basis functions and a maximum $N_\textnormal{split}$ of $\sim 3\times 10^4$. Bottom: Projection using fully adaptive grid, twice the number of image pixels as are source pixels. Both panels use curvature regularization, with the regularization strength optimized at each step.}
  \label{fig:varye}
\end{figure}

Here, we turn to examining the ability of the shapelets method to infer lens model parameters and compare the results against the well-established pixel-based approach.
The lens model is the same as in the previous section, but the source is now an elliptical Gaussian, placed in a cusp configuration, and additive Gaussian noise is applied, so that a peak S/N$\approx 41$ is achieved.
In Fig.~\ref{fig:varye}, we show a one-dimensional projection of the evidence as a function of ellipticity.
All other lens model parameters were fixed at their true values.
Because shapelets do not require a source grid and because the pixelization of the image plane is accounted for by integrating the shapelets in the source plane, the resulting $\chi^2$ curve is relatively smooth.
We find typical fluctuations of $\Delta\chi^2 <1$ using the shapelets-based method, while pixel-based methods can result in $\Delta\chi^2 \sim 10$.

\begin{figure}
  \centering
  \begin{tabular}{ c }
  \includegraphics[width=0.45\textwidth]{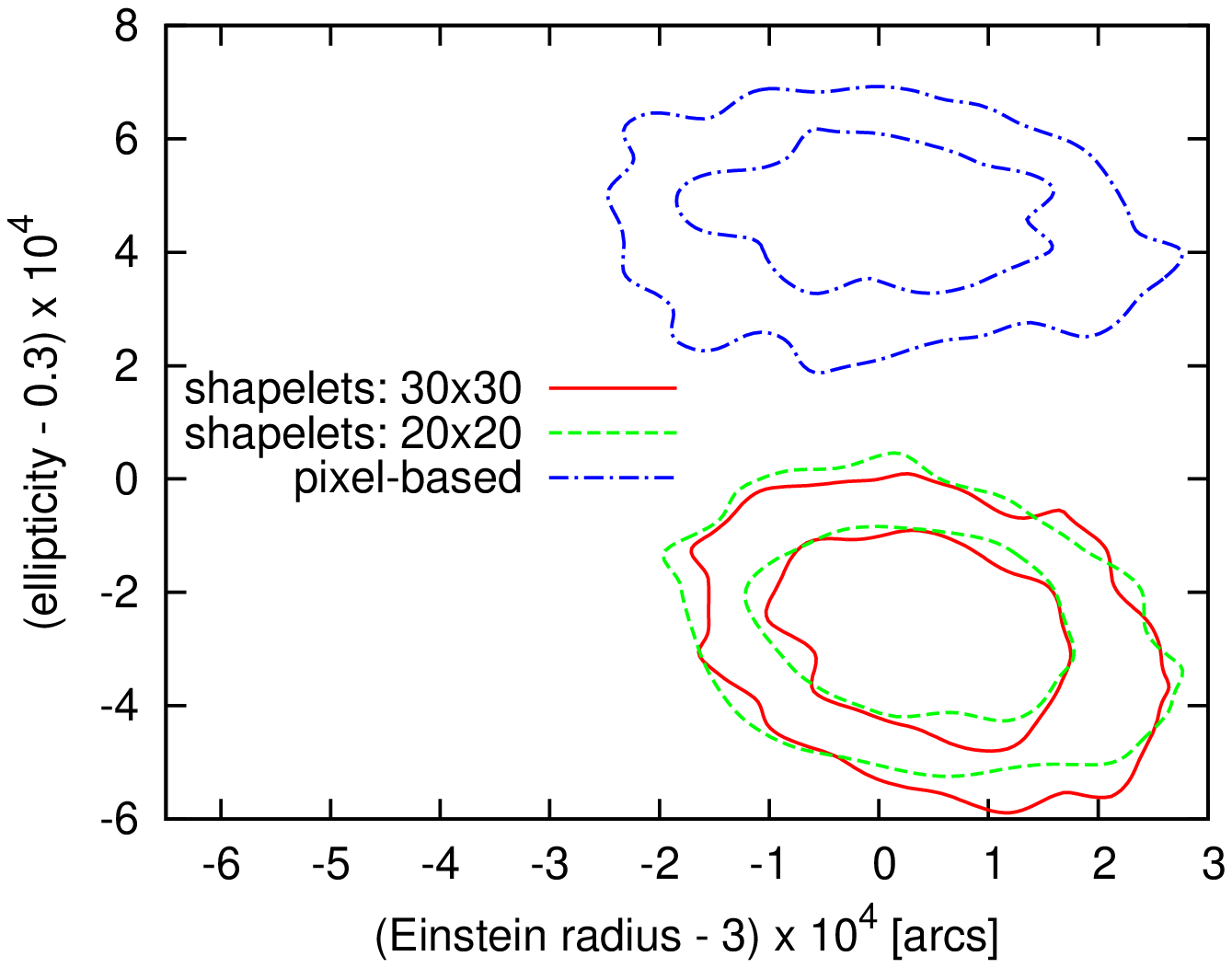} \\
  \includegraphics[width=0.45\textwidth]{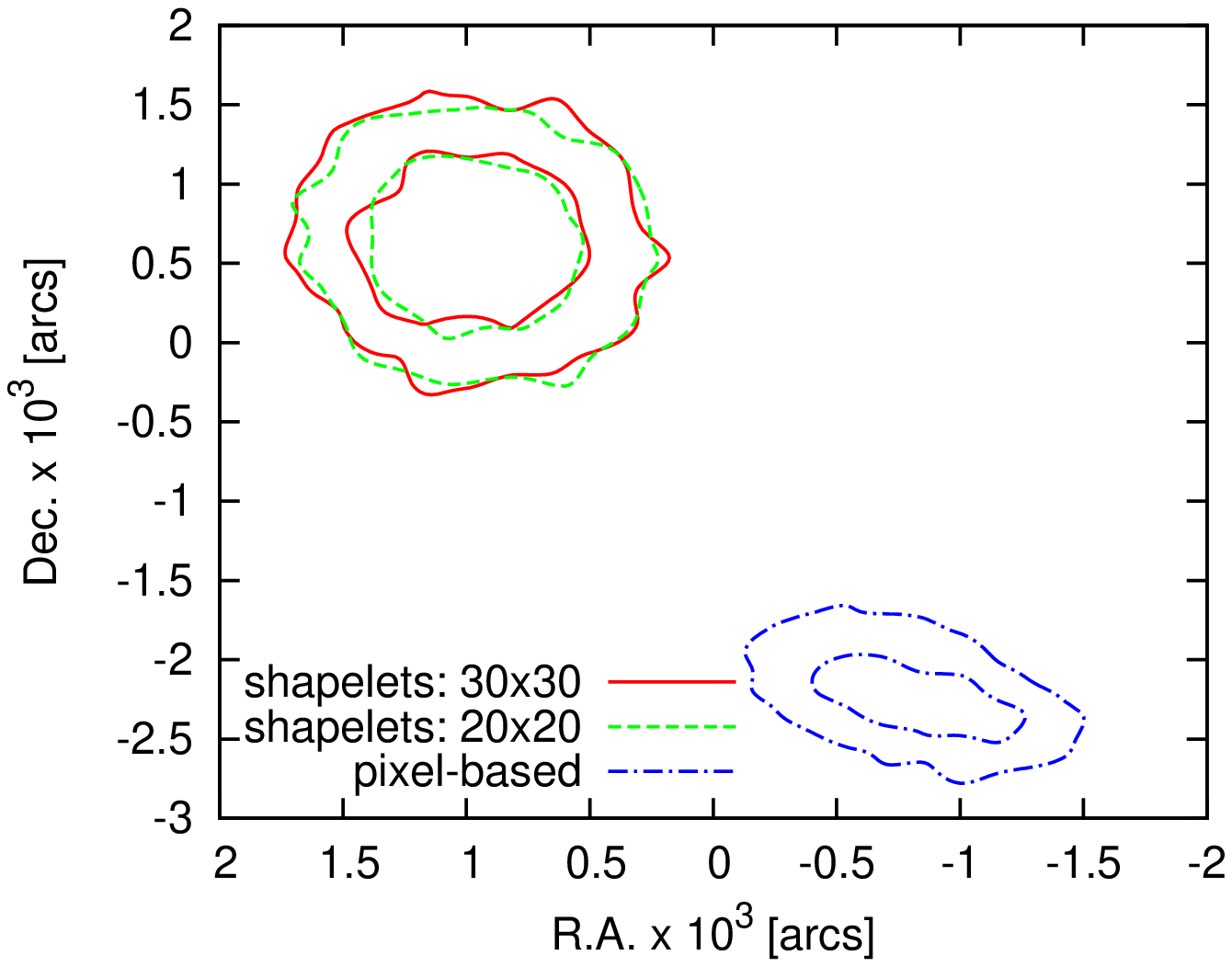}
  \end{tabular}
  \caption{\small
Joint posterior probability distributions for (top) Einstein radius and ellipticity and (bottom) right ascension and declination. For both panels, the true values of the lens model parameters have been subtracted and scaled, so that the origins of the coordinate systems coincide with the true lens model parameters. Because of the high peak S/N$\approx 300$, the uncertainties on the parameters are small, but (not unexpectedly) all source reconstruction methods show a clear bias on the estimate for all parameters, with the exception of the Einstein radius.}
  \label{fig:mcmc}
\end{figure}

Finally, we also perform a full lens model parameter space exploration, varying the Einstein radius, position, ellipticity, and position angle of the lens.
The source is still an elliptical Gaussian in a cusp configuration but, this time, the data have a peak S/N$\approx 300$ and an elliptical Gaussian PSF has been applied, with FWHMs of $0.236''$ and $0.165''$ along the major and minor axes, respectively, and a position angle of $-55^\circ$ EoN.
Joint posterior probability distributions for two pairs of lens model parameters are shown in Fig.~\ref{fig:mcmc}.
The shapelets-based method does a reasonable job of inferring the lens model parameters.
Except for the right ascension, all parameters are recovered at the 95\% confidence level.
The Einstein radius is the most robust of all the parameters, for all source reconstructions methods.
However, there is a clear bias in the positional parameters and the ellipticity.
For data of high S/N, such as these, an observed bias is not unexpected \citep[see, e.g,][]{tagore2014,nightingale2014}.
Interestingly, the biases in the shapelets and pixel-based approaches seem to be in opposite directions, but we have not explored this result fully.

\section{Conclusions}
\label{sec:conclusions}

We have developed an implementation of a shapelets-based, as opposed to pixel-based, approach for reconstructing resolved, gravitationally lensed images.
Expanding the source's surface brightness as a finite series of shapelets requires a choice of several parameters.
We have chosen to fix the center and scale of the shapelets using the data and lens model themselves.
Using mock data, \citet{birrershapelets} are able to accurately reconstruct their input source model and fit the lensed images.
We find that the source reconstruction can, for many lens configurations, lead to unphysical source reconstructions.
We therefore fix the maximum shapelets order and the maximum number of image pixel splittings $N_\textnormal{split}$ beforehand and regularize the source using a Bayesian framework, which stabilizes the reconstruction.
If there are not enough shapelet coefficients to accurately describe the source, model residuals will clearly indicate this.
We also find that for more extended sources, $N_\textnormal{split}$ does not play a significant role.
For smaller sources, $N_\textnormal{split}$ can be increased to test for robustness of the source reconstruction.

The advantages of the shapelets lie in the relatively small number of parameters needed to reproduce the data.
We find that even for complicated morphologies, $20\times 20=400$ source parameters are sufficient to capture most details.
On the other hand, creating the lensing operator requires evaluating every shapelet coefficient for every pixel, which  results in a dense, completely filled lensing operator matrix.
In practice, this does not result in a decrease in performance, because the recursion relations presented in \S\ref{sec:lo} allow for quick calculations as long as $N_\textnormal{split}$ is not unnecessarily large.
Moreover, a bottleneck in pixel-based algorithms occurs during optimization of the regularization strength.\footnote{Some authors omit this optimization and account for its effects in other ways.}
Because of the small number of source parameters, this step can be performed quickly, giving an unbiased estimator of the lens model being ranked.

We note, however, that given data exhibiting more complicated features than we have explored here (e.g., many starburst regions that are individually resolved or multiple, widely separated source galaxies), a large number of shapelets may be required, resulting in a decrease in performance.
A possible solution to this is to create multiple shapelet expansions centered at the appropriate positions, but we have not explored this possibility.




\appendix

\section{Integral approximation}
The $I_{0,0}$ integral cannot be calculated analytically.
It can be reduced to a single integral:
\begin{equation}
\begin{aligned}
I_{0,0} &= \frac{1}{\sqrt{2}}\int\limits_c^d\md x\merf\bigg(\frac{m x+b}{\sqrt{2}\beta}\bigg)\me^{-x^2/(2\beta^2)} \\
&= \frac{\beta}{m}\int\limits_{(m c+b)/(\sqrt{2}\beta)}^{(m d+b)/(\sqrt{2}\beta)}\md u\merf(u)\me^{-(\sqrt{2}\beta u-b)^2/(2\beta^2 m^2)}.
\label{eq:i00}
\end{aligned}
\end{equation}
By approximating the error function as the product of an exponential and a truncated power series expansion, the integral can be computed analytically.
Specifically, we use
\begin{equation}
\merf(u)\approx 1-\me^{d_1 u + d_2 u^2}(1+d_3 u + d_4 u^2 + d_5 u^3 + d_6 u^4).
\end{equation}
Optimizing the six free parameters, we find that
\begin{equation}
\begin{aligned}
&d_1 = -1.6093661701173445, \\
&d_2 = -0.9144569742603699, \\
&d_3 = 0.48105948656562003, \\
&d_4 = 0.3927744807707903, \\
&d_5 = 0.05250001720254402, \\
&\textnormal{and} \\
&d_6 = 0.036174940195734834.
\end{aligned}
\end{equation}

Although the approximation is valid only for positive arguments, negative arguments can be evaluated by realizing the error function is odd.
Over the entire domain of the function, we find a maximum absolute error of $2\times 10^{-6}$ and maximum relative error of $6\times 10^{-5}$.
Plugging this approximation into eq.~\ref{eq:i00}, an analytic solution can be found, although because of the large number of terms in the resulting expression, it cannot be sensibly reproduced here.

\bibliographystyle{mn2e}
\bibliography{refs}

\begin{thebibliography}{23}
\expandafter\ifx\csname natexlab\endcsname\relax\def\natexlab#1{#1}\fi

\bibitem[{{Birrer}, {Amara} \& {Refregier}(2015){Birrer}, {Amara}, \&
  {Refregier}}]{birrershapelets}
{Birrer} S., {Amara} A., {Refregier} A., 2015, ArXiv e-prints

\bibitem[{{Blandford}, {Surpi} \& {Kundi{\'c}}(2001){Blandford}, {Surpi}, \&
  {Kundi{\'c}}}]{blandfordEring}
{Blandford} R., {Surpi} G., {Kundi{\'c}} T., 2001, in Astronomical Society of
  the Pacific Conference Series, Vol. 237, Gravitational Lensing: Recent
  Progress and Future Goals, {Brainerd} T.~G., {Kochanek} C.~S., eds., p.~65

\bibitem[{{Brewer} \& {Lewis}(2006)}]{citebrewer2006}
{Brewer} B.~J., {Lewis} G.~F., 2006, \apj, 637, 608

\bibitem[{{Dye} {et~al}\mbox{.}(2013){Dye}, {Negrello}, {Hopwood},
  {Nightingale}, {Bussmann}, {Amber}, {Bourne}, {Cooray}, {Dariush}, {Dunne},
  {Eales}, {Gonzalez-Nuevo}, {Ibar}, {Ivison}, {Maddox}, {Valiante}, \&
  {Smith}}]{sourceplanescience2}
{Dye} S. {et~al.}, 2013, ArXiv e-prints

\bibitem[{{Dye} \& {Warren}(2005)}]{dyegrid}
{Dye} S., {Warren} S.~J., 2005, \apj, 623, 31

\bibitem[{{Keeton}(2001)}]{citelensmodel}
{Keeton} C.~R., 2001, arXiv:astro-ph/0102340

\bibitem[{{Kochanek}, {Keeton} \& {McLeod}(2001){Kochanek}, {Keeton}, \&
  {McLeod}}]{kochanekEring}
{Kochanek} C.~S., {Keeton} C.~R., {McLeod} B.~A., 2001, \apj, 547, 50

\bibitem[{{Koopmans}(2005)}]{koopmans:cdmsubstructure}
{Koopmans} L.~V.~E., 2005, \mnras, 363, 1136

\bibitem[{{Massey} \& {Refregier}(2005)}]{massey_shapeletsII}
{Massey} R., {Refregier} A., 2005, \mnras, 363, 197

\bibitem[{{Nightingale} \& {Dye}(2014)}]{nightingale2014}
{Nightingale} J., {Dye} S., 2014, ArXiv e-prints

\bibitem[{{Refregier}(2003)}]{refregier_shapeletsI}
{Refregier} A., 2003, \mnras, 338, 35

\bibitem[{{Rybak} {et~al}\mbox{.}(2015){Rybak}, {McKean}, {Vegetti},
  {Andreani}, \& {White}}]{vegettisdp81}
{Rybak} M., {McKean} J.~P., {Vegetti} S., {Andreani} P., {White} S.~D.~M.,
  2015, ArXiv e-prints

\bibitem[{{Schneider}, {Kochanek} \& {Wambsganss}(2006){Schneider}, {Kochanek},
  \& {Wambsganss}}]{saasfee}
{Schneider} P., {Kochanek} C.~S., {Wambsganss} J., 2006, {Gravitational
  Lensing: Strong, Weak and Micro}. Springer

\bibitem[{{Sharon} {et~al}\mbox{.}(2012){Sharon}, {Gladders}, {Rigby}, {Wuyts},
  {Koester}, {Bayliss}, \& {Barrientos}}]{sourceplanescience}
{Sharon} K., {Gladders} M.~D., {Rigby} J.~R., {Wuyts} E., {Koester} B.~P.,
  {Bayliss} M.~B., {Barrientos} L.~F., 2012, \apj, 746, 161

\bibitem[{{Suyu} \& {Halkola}(2010)}]{suyu:substr}
{Suyu} S.~H., {Halkola} A., 2010, \aap, 524, A94

\bibitem[{{Suyu} {et~al}\mbox{.}(2006){Suyu}, {Marshall}, {Hobson}, \&
  {Blandford}}]{suyureg}
{Suyu} S.~H., {Marshall} P.~J., {Hobson} M.~P., {Blandford} R.~D., 2006,
  \mnras, 371, 983

\bibitem[{{Tagore} \& {Keeton}(2014)}]{tagore2014}
{Tagore} A.~S., {Keeton} C.~R., 2014, \mnras, 445, 694

\bibitem[{{Vegetti}, {Czoske} \& {Koopmans}(2010){Vegetti}, {Czoske}, \&
  {Koopmans}}]{vegetticlone}
{Vegetti} S., {Czoske} O., {Koopmans} L.~V.~E., 2010, \mnras, 407, 225

\bibitem[{{Vegetti} \& {Koopmans}(2009)}]{vegettigrid}
{Vegetti} S., {Koopmans} L.~V.~E., 2009, \mnras, 392, 945

\bibitem[{{Vegetti} {et~al}\mbox{.}(2010){Vegetti}, {Koopmans}, {Bolton},
  {Treu}, \& {Gavazzi}}]{vegetti:dark}
{Vegetti} S., {Koopmans} L.~V.~E., {Bolton} A., {Treu} T., {Gavazzi} R., 2010,
  \mnras, 408, 1969

\bibitem[{{Vegetti} {et~al}\mbox{.}(2012){Vegetti}, {Lagattuta}, {McKean},
  {Auger}, {Fassnacht}, \& {Koopmans}}]{vegetti:dark2}
{Vegetti} S., {Lagattuta} D.~J., {McKean} J.~P., {Auger} M.~W., {Fassnacht}
  C.~D., {Koopmans} L.~V.~E., 2012, \nat, 481, 341

\bibitem[{{Wallington}, {Kochanek} \& {Narayan}(1996){Wallington}, {Kochanek},
  \& {Narayan}}]{wallington1996}
{Wallington} S., {Kochanek} C.~S., {Narayan} R., 1996, \apj, 465, 64

\bibitem[{{Warren} \& {Dye}(2003)}]{citewarrendyesemilinear}
{Warren} S.~J., {Dye} S., 2003, \apj, 590, 673

\end{thebibliography}

\clearpage

\end{document}